\documentclass[12pt,a4paper,aps,amsmath,byrevtex]{revtex4}
\usepackage{graphicx}
\usepackage{wasysym}
\newcommand{\bm}{\boldmath}
\newcommand{\D}{\displaystyle}
\newcommand{\vek}[1]{\mbox{\bm ${#1}$}}
\begin{document}
\begin{center}
{\bf Corrected Draft 08 August 2006}
\end{center}
\thispagestyle{empty}
\renewcommand{\theequation}{\arabic{section}.\arabic{equation}}
\vspace{3cm}
\begin{center}{\large\bf The self-energy of the uniform electron gas in the second order of exchange}\\
\vspace{2cm}
{\sc P. Ziesche} \\
\vspace{1cm}
{Max-Planck-Institut f\"ur Physik komplexer Systeme \\
N\"othnitzer Str. 38, D-01187 Dresden, Germany \\
pz@pks.mpg.de}\\
\vspace{1cm}
\date{\today}
\vspace{1cm}
\end{center}


\begin{abstract}
\noindent
The on-shell self-energy of the homogeneous electron gas in second order of exchange, $\Sigma_{2{\rm x}}=
{\rm Re}\ \Sigma_{2{\rm x}}(k_{\rm F},k_{\rm F}^2/2)$, is given by a 
certain integral. This integral is treated here in a similar way as Onsager, Mittag, and Stephen [Ann. Physik 
(Leipzig) {\bf 18}, 71 (1966)] have obtained their famous analytical expression $e_{2{\rm x}}=\frac{1}{6}\ln 2-
\frac{3}{4}\frac{\zeta(3)}{\pi^2}$ (in atomic units) for the correlation energy in second order of exchange. Here it is shown
that the result for the corresponding on-shell self-energy is $\Sigma_{2{\rm x}}=e_{2{\rm x}}$.  
The off-shell self-energy $\Sigma_{2{\rm x}}(k,\omega)$ correctly 
yields $2e_{2{\rm x}}$ (the potential component of $e_{2{\rm x}}$) through the
Galitskii-Migdal formula. The quantities $e_{2{\rm x}}$ and 
$\Sigma_{2{\rm x}}$ appear in the high-density limit of the Hugenholtz-van Hove (Luttinger-Ward) theorem.
\end{abstract}
\maketitle
\newpage
\section{Introduction}
\noindent
Although not present in the Periodic Table the homogeneous electron gas (HEG) is still an important 
model system for electronic structure theory, cf. e.g. \cite{Tos}. In its spin-unpolarized version 
the HEG ground state is characterized by only one parameter $r_s$, such that a sphere with the radius 
$r_s$ contains {\it on average} one electron \cite{Zie1}. It determines the Fermi wave number as 
$k_{\rm F}=1/(\alpha r_s)$ with $\alpha =(4/(9\pi))^{1/3}$ and it measures simultaneously the 
interaction strength and the density such that high density corresponds to weak interaction and hence 
weak correlation. For recent papers on this limit cf. \cite{Zie2,Zie3,Mui}. Naively one should 
expect that in this weak-correlation limit the Coulomb repulsion $\alpha r_s/r$ (where lengths and 
energies are measured in units of $k_{\rm F}^{-1}$ and $k_{\rm F}^2$, respectively) can be treated as 
perturbation. But in the early theory of the HEG, Heisenberg \cite{Hei} has shown, that ordinary 
perturbation theory does not apply. With $e_0$ being the energy per particle of the ideal Fermi gas and 
$e_{\rm x}$ being the exchange energy in lowest (1st) order, the total energy  $e=e_0+e_{\rm x}+e_{\rm c}$ 
defines the correlation energy $e_{\rm c}=e_2+e_3+\cdots$. In 2nd order, there is a 
direct (d) term 
$e_{2{\rm d}}$ and an exchange (x) term $e_{2{\rm x}}$, so that $e_2=e_{2{\rm d}}+e_{2{\rm x}}$. Whereas 
$e_{2{\rm x}}/(\alpha r_s)^2$ is a pure finite number $b_{\rm x}$,
the direct term $e_{2{\rm d}}$ logarithmically diverges along the Fermi surface (i.e. for 
vanishing transition momenta $q$): $e_{2{\rm d}}\to\ln q$ for $q\to 0$. This failure of perturbation theory has been
repaired by Macke \cite{Ma} with an appropriate partial summation of higher-order terms up to infinite order.
The result of this (ring-diagram) summation for the correlation energy in its weak-correlation limit is 
$e_{\rm c}/(\alpha r_s)^2=a\ln r_s +
\cdots$ with $a=(1-\ln 2)/\pi^2\approx 0.031091$. This has been confirmed later by Gell-Mann and Brueckner 
\cite{GB}, who in addition to the logarithmic term numerically calculated contributions to the next (constant, i.e. 
not depending on $r_s$) term $b$, namely $b_{\rm r}$ and $b_{\rm d}$ arising from the ring-diagram summation and from 
$e_{2{\rm d}}$, respectively: $b_{\rm r}\approx a\ln \frac{\alpha}{\pi}-0.001656\approx -0.057514$ and 
$b_{\rm d}\approx -0.013586$. The total constant term is $b=b_{\rm r}+b_{\rm d}+b_{\rm x}\approx -0.046921$, where 
the exchange term $b_{\rm x}=\frac{1}{6}\ln 2-\frac{3}{4}\frac{\zeta(3)}{\pi^2}\approx +0.0241792$ has been 
{\it analytically} calculated by Onsager, Mittag, and Stephen in a very tricky way \cite{Ons}. Although the integrand
is rather simple, the Pauli principle causes complicated boundary conditions, which need a sophisticated 
treatment through several substitutions of the integral variables. \\

\noindent
Here their method is used to calculate the analog term of the self-energy on the energy shell, namely
$\Sigma_{2{\rm x}}={\rm Re}\ \Sigma_{2{\rm x}}(1,1/2)$ [with $k$ measured in units of $k_{\rm F}$ and $\omega$ 
measured in units of $k_{\rm F}^2$]. The more general quantity $\Sigma_{2{\rm x}}(k,\omega)$ appears (i) in a recent 
study of the spectral
moments of the HEG \cite{Vog}, (ii) in the context of self-consistent $GW$ calculations \cite{Shir}, and 
within a project 'spectral functions for a hydrogen plasma' \cite{Fort}. For $k=1$, $\omega=\mu$ it 
appears in the Hugenholtz-van Hove (Luttinger-Ward) identity \cite{Hug}, which 
relates the chemical potential $\mu$ to the self-energy $\Sigma(k,\omega)$ 
according to $\mu -\mu_0=\Sigma(1,\mu)$. The chemical potential follows from the total energy according to 
$\mu=(\frac{5}{3}-\frac{1}{3}r_s\frac{d}{dr_s})e$. In the weak-correlation limit $r_s\to 0$ the total energy 
$e=e_0+e_{\rm x}+e_{\rm c}$ and the chemical potential $\mu=\mu_0+\mu_{\rm x}+\mu_{\rm c}$ are given by 
\begin{eqnarray}\label{aa1}
\ e_0=\frac{3}{10},  \quad \qquad e_{\rm x}=-\frac{3}{4}\frac{\alpha r_s}{\pi}, 
\quad \qquad e_{\rm c}=(\alpha r_s)^2[a\ln r_s+b+O(r_s)], \nonumber \\
\quad \mu_0=\frac{1}{2}, \quad \mu_{\rm x}=-\frac{\alpha r_s}{\pi}, \; \quad 
\mu_{\rm c}=(\alpha r_s)^2\left[a\ln r_s+\left(-\frac{1}{3}a+b\right)+O(r_s)\right].  
\end{eqnarray}
Because of the above mentioned identity the self-energy $\Sigma(1,\mu)$ must have a corresponding behavior. 
The exchange in lowest order yields $\Sigma_{\rm x}(1)=-\frac{\alpha r_s}{\pi}$ or $\mu_{\rm x}=\Sigma_{\rm x}(1)$, 
exactly in agreement with 
the mentioned identity. To obtain also the logarithmic term of $\Sigma(1,1/2)$ the 
ring-diagram summation has to be done for the self-energy \cite{Zie4}. To the next term beyond the logarithmic term
contributes the 2nd-order exchange self-energy  
$\Sigma_{2{\rm x}}={\rm Re}\ \Sigma_{2{\rm {\rm x}}}(1,1/2)$. Just this term is calculated here 
using the tricky method of Onsager, Mittag, and Stephen \cite{Ons} {\it mutatis mutandi}. 
The Feynman diagrams of $e_{2{\rm x}}$ and $\Sigma_{2{\rm x}}$ are given in Figs. 1 and 2. 
As shown in the Appendix, from the Feynman diagram rules it follows $\Sigma_{2{\rm x}}= 
-\frac{(\alpha r_s)^2}{4\pi^4}(X_1+X_2)$, where $X_{1,2}$ mean the integrals defined in Eqs. (\ref{A10}) and 
(\ref{A11}). They are calculated in the following sections. The final results are $X_1=-\pi^4\left[\frac{4}{3}\ln 2-
5\frac{\zeta(3)}{\pi^2}\right]$, 
$X_2=\pi^4\left[\frac{2}{3}\ln 2-2\frac{\zeta(3)}{\pi^2}\right]$, thus $\Sigma_{2{\rm x}}=
(\alpha r_s)^2\left[\frac{1}{6}\ln 2-\frac{3}{4}\frac{\zeta(3)}{\pi^2}\right]=e_{2{\rm x}}$, 
thus $\mu_{2{\rm x}}=\Sigma_{2{\rm x}}$. This relation appears in the weak-correlation limit of  
the above mentioned Hugenholtz-van Hove theorem \cite{Zie4,Hug}. This paper is a 
contribution to the mathematics of the weakly-correlated (high-density) HEG.

\section{The Integral $X_1$} 
\setcounter{equation}{0}

\noindent
In Eq. (\ref{A10}), new variables ${\vek q}'=({\vek q}_1+{\vek q}_2)/2$,
${\mbox{\bm $p$}'}=({\mbox{\bm $q$}}_1-{\mbox{\bm $q$}}_2)/2$, and 
${\mbox{\bm $s$}'}={\mbox{\bm $e$}}+({\mbox{\bm $q$}}_1+{\mbox{\bm $q$}}_2)/2$ lead to  
\begin{eqnarray}\label{b1}
{\mbox{\bm $q$}}_1={\vek q}'+{\vek p}', \quad \quad \quad ({\mbox{\bm $e+q$}}_1)^2&=&({\vek s}'+{\vek p}')^2, \quad
({\mbox{\bm $e+q$}}_2)^2=({\vek s}'-{\vek p}')^2, \quad \nonumber \\
{\vek q}_2={\vek q}'-{\vek p}', \quad
({\mbox{\bm $e$}}+{\mbox{\bm $q$}}_1+{\mbox{\bm $q$}}_2)^2&=&({\vek s}'+{\vek q}')^2, \quad \quad \quad \quad \; 
{\mbox{\bm $e$}}^2=({\vek s}'-{\vek q}')^2.
\end{eqnarray}
Therefore (with $\Theta(x)$ = Heaviside step function) 
\begin{eqnarray}
X_1=\int d^3q'd^3p'\ \frac{2^3}{q'^2-p'^2}\  \frac{1}
{({\mbox{\bm $p$}'}+{\mbox{\bm $q$}'})^2({\mbox{\bm $p$}'}-{\mbox{\bm $q$}'})^2}\
\times \nonumber
\end{eqnarray}
\begin{eqnarray}\label{b2} 
\times\int\frac{d^2s'}{\pi}\ \delta(({\mbox{\bm $s$}'}-{\mbox{\bm $q$}'})^2-1)
\Theta[1-({\mbox{\bm $s$}'}+{\mbox{\bm $q$}'})^2]  
\Theta[({\mbox{\bm $s$}'}+{\mbox{\bm $p$}'})^2-1]
\Theta[({\mbox{\bm $s$}'}-{\mbox{\bm $p$}'})^2-1].   
\end{eqnarray}
The next step scales ${\vek q}'$, ${\vek p}'$, ${\vek s}'$ with
$\lambda=1/{\sqrt {1-s'^2}}$ according to $ {\mbox{\bm $q$}}=\lambda{\mbox{\bm $q$}'}$, 
${\mbox{\bm $p$}}=\lambda{\mbox{\bm $p$'}}$, ${\mbox{\bm $s$}}=\lambda{\mbox{\bm $s$'}}$ 
with the consequences   
\begin{eqnarray}\label{b3}
1-s'^2=\frac{1}{1+s^2}, \quad d^2s'=\frac{d^2s}{(1+s^2)^2}, \nonumber \\
\delta(({\mbox{\bm $s$}}'-{\mbox{\bm $q$}}')^2-1)=(1+s^2)\ \delta(q^2-2{\mbox{\bm $s$}}{\mbox{\bm $q$}}-1), 
\nonumber \\
\pm 2{\mbox{\bm $s$}}{\mbox{\bm $p$}}+p^2>1, \quad 2{\mbox{\bm $s$}}{\mbox{\bm $q$}}+q^2<1, \quad 
-2{\mbox{\bm $s$}}{\mbox{\bm $q$}}+q^2=1,
\end{eqnarray}
from which follow $p\geq 1$ and $q\leq 1$ (what makes the energy denominator $q^2-p^2$ negative) and 
${\vek s} {\vek q}<0$, $s\geq \alpha$. Thus 
\begin{eqnarray}\label{b4}
X_1=-\int\limits_0^1  dq\int\limits_1^\infty  dp\; \frac{16\ \pi}{p^2-q^2}
\int\limits_{-1}^{+1}\frac{dx}{2}\ \frac{8\ q^2p^2}{(p^2+q^2)^2-(2{\mbox{\bm $q$}}{\mbox{\bm $p$}})^2} \times  
\nonumber \\
\times\int\limits_{s\geq \alpha,\ \cos \varphi_q<0}\frac{d^2s}{1+s^2}\ \delta(q^2-1-2{\mbox{\bm $s$}}{\mbox{\bm $q$}})\ 
\Theta(p^2-1+2{\mbox{\bm $s$}}{\mbox{\bm $p$}})\ \Theta(p^2-1-2{\mbox{\bm $s$}}{\mbox{\bm $p$}}). 
\end{eqnarray}
Here and in the following of Sec. II the abbreviations
\begin{eqnarray}\label{b5}
\alpha=\frac{1-q^2}{2q}\geq 0,\quad \beta=\frac{p^2-1}{2p}\geq 0, \quad a=\frac{q^2+p^2}{2qp}\geq 1, 
\nonumber \\
\frac{t}{s}=\cos\varphi_p=\cos\varangle ({\vek s},{\vek p}), \quad
x=\cos\varphi=\cos\varangle ({\vek q},{\vek p}),\quad 
y=\cos\varphi_q=\cos\varangle ({\vek q},{\vek s})
\end{eqnarray}
are used:
\begin{eqnarray}\label{b6}
X_1=-\int\limits_0^1  dq\int\limits_1^\infty  dp\; \frac{16\ \pi}{p^2-q^2}
\int\limits_{-1}^{+1} \frac{dx}{a^2-x^2} 
\int\limits_{s\geq \alpha,\ \cos \varphi_q<0 }\frac{sds\ d\varphi_q}{1+s^2}\ \frac{1}{2sq}\ 
\delta\left(\frac{\alpha}{s}+y\right)\ \Theta(\beta-|t|)\ . 
\end{eqnarray}
$t$ is introduced to replace $s$ after having done the $\varphi_q$ integration. To this purpose
the relation between $t$ and $s$ is needed. It follows from 
$\varphi_q+\varphi_p=\varphi$ or $\varphi_p=\varphi-\varphi_q$, therefore 
$\cos \varphi_p=\cos (\varphi-\varphi_q)= 
\cos \varphi\cos \varphi_q+\sin \varphi \sin \varphi_q$, what is in terms of $t/s$, $x$, and 
$y$ (the latter quantity equals $-\alpha/s$ because of the delta function):
\begin{equation}\label{b7}
\frac{t}{s}=x\left(-\frac{\alpha}{s}\right)\pm\sqrt {1-x^2}\sqrt {1-\left(\frac{\alpha}{s}\right)^2}
\quad {\rm or} \quad (t+\alpha x)^2=(1-x^2)(s^2-\alpha^2).
\end{equation}
This has the consequences
\begin{eqnarray}\label{b8}
s^2-\alpha^2=\frac{(t+\alpha x)^2}{1-x^2}\ , \\
\frac{s ds}{1+s^2}=\frac{(t+\alpha x)dt}{(t+\alpha x)^2+(1+\alpha^2)(1-x^2)}\ .
\end{eqnarray}
With the help of Eq. (\ref{b8}) the $\varphi_q$ integration yields a function depending on $s$
(respectively on $t$). Note that $\cos \varphi_q=-\frac{\alpha}{s}$ has two solutions $\varphi_{q,1}$ and 
$\varphi_{q,2}$ with $|\sin \varphi_{q,1}|=|\sin \varphi_{q,2}|=\sqrt {1-(\frac{\alpha}{s})^2}$ :
\begin{eqnarray}\label{b10}
\int\limits_0^{2\pi}\frac{d\varphi_q}{s}\ \left [\delta(-\sin\varphi_{q,1} \cdot (\varphi_q-\varphi_{q,1}))+
\delta(-\sin\varphi_{q,2} \cdot (\varphi_q-\varphi_{q,2}))\right] \nonumber \\
=\int\limits_0^{2\pi}\frac{d\varphi_q}{s}\ \frac{\delta(\varphi_q-\varphi_{q,1})+\delta (\varphi_q-\varphi_{q,2})}{\sqrt {1-(\frac{\alpha}{s})^2}}=
\frac{2}{\sqrt {s^2-\alpha^2}}=2\ \frac{\sqrt {1-x^2}}{|t+\alpha x|}\ .
\end{eqnarray} 
Thus 
\begin{eqnarray}\label{b11}
\int\limits_{s\geq \alpha,\ \cos \varphi_q }\frac{sds\ d\varphi_q}{1+s^2}\ \frac{1}{2sq}\ 
\delta\left(\frac{\alpha}{s}+y\right)\ \Theta(\beta-|t|)&=&
\frac{\sqrt {1-x^2}}{q}\int\limits_{t_0}^{t_1}\frac{{\rm sign}(t+\alpha x)\ dt}{(t+\alpha x)^2+(1+\alpha^2)(1-x^2)} 
\nonumber \\ 
&=&\left.\frac{2}{1+q^2}\arctan \frac{t+\alpha x}{\sqrt {(1+\alpha^2)(1-x^2)}}\right |_{t_0}^{t_1}.
\end{eqnarray}
$|t|\leq\beta$ contains the upper 
limit of the $t$-integration as $t\leq\beta=:t_1$. What concerns its lower limit $t_0$, the relation 
$t\geq -\beta$
competes with $t\geq -\alpha x$, as it follows from Eq. (\ref{b7}) for the lower limit $\alpha$ of the
$s$-integration. This means for the ranges of the $t$- and $x$-integrations, one has in the 
(vertical) stripe $q=0\cdots1, 
p=1\cdots\infty$ of the $q$-$p$-plane to distinguish between the two regions, cf. Fig. 3: \\
(a) region A with $qp\geq 1$ or $\alpha\leq \beta$ or $-\alpha x\geq -\alpha\geq -\beta$, hence $t_0=-\alpha x$, and \\
(b) region B with $qp\leq 1$ or $\alpha\geq \beta$ or $-\alpha\leq -\beta$. \\
In the case (b) one has again to distinguish between \\
(i) $x=-\beta/\alpha \cdots +\beta/\alpha$, with the consequence $t_0=-\alpha x$ and \\
(ii) $x=\beta/\alpha\cdots 1$, with the consequence $t_0=-\beta$. \\
Thus it is
\begin{eqnarray}\label{b12}
\int\limits_{-1}^{+1}dx\int\limits_{t_0}^{+\beta}dt=\left\{
\begin{array} {ll}
\int\limits_{-1}^{+1}dx\int\limits_{-\alpha x}^{+\beta}dt \; \quad \qquad \qquad \qquad {\rm for} 
\quad qp\geq 1\; {\rm or}\; \alpha\leq \beta , \\
\\
\int\limits_{-\beta/\alpha}^{+\beta/\alpha}dx\int\limits_{-\alpha x}^{+\beta}dt+
\int\limits_{+\beta/\alpha}^{+1}dx\int\limits_{-\beta}^{+\beta}dt \quad {\rm for} \quad qp\leq 1\; 
{\rm or}\; \alpha\geq \beta .
\end{array}
\right.
\end{eqnarray} 
With the abbreviations (\ref{b5}) and
\begin{equation}\label{b13}
f(t,x)= \frac{2}{a^2-x^2}\ \arctan \frac{t+\alpha x}{\sqrt {(1+\alpha^2)(1-x^2)}}\ ,
\end{equation}
Eq. (\ref{b6}) can be written as (note that $f(-\alpha x,x)=0$)
\begin{eqnarray}\label{b14}
X_1=-\int\limits_0^1 dq \int\limits_1^\infty dp\ \frac{16\ \pi}{(p^2-q^2)}\ \frac{1}{(1+q^2)}\ 
\left\{\Theta(qp-1) \int\limits_{-1}^{+1}dx\ f(\beta,x)\ \right. \nonumber \\  
+\left.\Theta(1-qp)\left[\int\limits_{-\beta/\alpha}^{+\beta/\alpha}dx\ f(\beta,x) 
+ \int\limits_{+\beta/\alpha}^{+1}dx\ [f(\beta,x)-f(-\beta,x)]\right]\right\}\ . 
\end{eqnarray}
With 
\begin{equation}\label{b15}
\int\limits_{+\beta/\alpha}^{+1}dx\ (-1)f(-\beta,x)=\int\limits_{+\beta/\alpha}^{+1}dx\ f(\beta,-x)=
\int\limits_{-1}^{-\beta/\alpha}dx\ f(\beta,x)
\end{equation}
the terms for $qp\leq 1$ can be comprised as $\int\limits_{-1}^{+1}dx\ f(\beta,x)$. Therefore
\begin{equation}\label{b16}
X_1=-\int\limits_{0}^{1}dq\int\limits_{1}^{\infty}dp\int\limits_{-1}^{+1}dx\
\frac{16\ \pi}{p^2-q^2}\ \frac{1}{1+q^2}\ f(\beta,x)\ .
\end{equation}
The final substitution $p=1/k$ transforms the region of the last two integrations from the
(vertical) stripe                                                                   
$q=0\cdots 1,p=1\cdots \infty$ to the more simple unit square $q=0\cdots 1,k=0\cdots 1 $.
With the abbrevations
\begin{equation}\label{b17}
\alpha=\frac{1-q^2}{2q}\geq 0,\quad \beta=\frac{1-k^2}{2k}\geq 0,\quad a=\frac{1+q^2k^2}{2qk}\geq 1
\end{equation}
it is:
\begin{equation}\label{b18}
X_1=-\int\limits_{0}^{1}dq\int\limits_{0}^{1}dk\int\limits_{-1}^{+1}dx\
\frac{16\ \pi}{1-q^2k^2}\ \frac{1}{1+q^2}\ f(\beta,x)\approx -30.70598\cdots\ . 
\end{equation}
The coefficients $\alpha$,  $\beta$, and $a$ of Eq. (\ref{b17}) 
make the integrand of (\ref{b18}) functions of $q$ and $k$. Mathematica5.2 \cite{math} yields the given figure.  
It seems to hold
\begin{equation}\label{b19}
X_1=-\pi^4\left[\frac{4}{3} \ln 2- 5\frac{ \zeta(3)}{\pi^2}\right]\approx -30.705985239248893\cdots
\end{equation}
How to derive this analytically ? Is this possible with the method of ref. \cite{Gla2} ? \\

\section{The Integral $X_2$}
\setcounter{equation}{0}

\noindent
The whole procedure of Sec. II is repeated 
here step by step. In Eq. (\ref{A11}), new variables ${\mbox{\bm $q$}'}=({\mbox{\bm $q$}}_1+{\mbox{\bm $q$}}_2)/2$,
${\mbox{\bm $p$}'}=({\mbox{\bm $q$}}_1-{\mbox{\bm $q$}}_2)/2$, and
${\mbox{\bm $s$}'}={\mbox{\bm $e$}}+({\mbox{\bm $q$}}_1+{\mbox{\bm $q$}}_2)/2$ lead to
\begin{eqnarray}\label{c1}
{\vek q}_1={\vek q}'+{\vek p}', \quad \quad \quad ({\mbox{\bm $e+q$}}_1)^2&=&({\vek s}'+{\vek p}')^2, \quad
({\mbox{\bm $e+q$}}_2)^2=({\vek s}'-{\vek p}')^2, \quad \nonumber \\
{\vek q}_2={\vek q}'-{\vek p}', \quad ({\mbox{\bm $e$}}+{\mbox{\bm $q$}}_1+{\mbox{\bm $q$}}_2)^2&=&({\vek s}'+
{\vek q}')^2, \quad \quad \quad \quad \; {\mbox{\bm $e$}}^2=({\vek s}'-{\vek q}')^2.
\end{eqnarray}
Therefore
\begin{eqnarray}
X_2=\int d^3q'd^3p'\ \frac{2^3}{q'^2-p'^2}\ 
\frac{1}{({\mbox{\bm $q$}}'+{\mbox{\bm $p$}'})^2({\mbox{\bm $q$}'}-{\mbox{\bm $p$}'})^2}
\ \times \nonumber
\end{eqnarray}
\begin{eqnarray}\label{c2}
\times\int\frac{d^2s'}{\pi}\ \delta(({\vek s}'-{\vek q}')^2-1)
\Theta[({\mbox{\bm $s$}'}+{\mbox{\bm $q$}'})^2-1]
\Theta[1-({\mbox{\bm $s$}'}+{\mbox{\bm $p$}'})^2]
\Theta[1-({\mbox{\bm $s$}'}-{\mbox{\bm $p$}'})^2]\ . 
\end{eqnarray}
The next step scales ${\mbox{\bm $q$}'}$, ${\mbox{\bm $p$}'}$, ${\mbox{\bm $s$}'}$ with
$\lambda=1/{\sqrt {1-s'^2}}$ according to $ {\mbox{\bm $q$}}=\lambda{\mbox{\bm $q$}'}$,
${\mbox{\bm $p$}}=\lambda{\mbox{\bm $p$'}}$, ${\mbox{\bm $s$}}=\lambda{\mbox{\bm $s$'}}$
with the consequences
\begin{eqnarray}\label{c3}
1-s'^2=\frac{1}{1+s^2}, \quad d^2s'=\frac{d^2s}{(1+s^2)^2}, \nonumber \\
\delta(({\mbox{\bm $s$}}'-{\mbox{\bm $q$}}')^2-1)=(1+s^2)\ \delta(q^2-2{\mbox{\bm $s$}}{\mbox{\bm $q$}}
-1), \nonumber \\
\pm 2{\mbox{\bm $s$}}{\mbox{\bm $p$}}+p^2<1, \quad 2{\mbox{\bm $s$}}{\mbox{\bm $q$}}+q^2>1, \quad
-2{\mbox{\bm $s$}}{\mbox{\bm $q$}}+q^2=1,
\end{eqnarray}
from which follow $q\geq 1$ and $p\leq 1$ (what makes the energy denominator $q^2-p^2$ positive) and 
${\vek s}{\vek q}>0$, $s\geq\bar\alpha$ . Thus
\begin{eqnarray}\label{c4}
X_2=\int\limits_1^\infty  dq\int\limits_0^1  dp\; \frac{16\ \pi}{q^2-p^2}
\int\limits_{-1}^{+1}\frac{dx}{2}\ \frac{8\ q^2p^2}{(p^2+q^2)^2-(2{\vek q}{\vek p)^2}} \times  \nonumber \\
\times\int\limits_{s\geq \bar\alpha,\ \cos \varphi_q>0}\frac{d^2s}{1+s^2}\ \delta(q^2-1-2{\vek s}{\vek q})
\Theta(1-p^2-2{\mbox{\bm $s$}}{\mbox{\bm $p$}})\Theta(1-p^2+2{\mbox{\bm $s$}}{\mbox{\bm $p$}}).
\end{eqnarray}
Here and in the following of Sec. III the abbreviations
\begin{eqnarray}\label{c5}
{\bar\alpha}=\frac{q^2-1}{2q}\geq 0,\quad {\bar\beta}=\frac{1-p^2}{2p}\geq 0, \quad 
\bar a=\frac{q^2+p^2}{2qp}\geq 1, \nonumber \\
\frac{t}{s}=\cos\varphi_p=\cos\varangle ({\vek s},{\vek p}), \quad
x=\cos\varphi=\cos\varangle ({\vek q},{\vek p}),\quad
y=\cos\varphi_q=\cos\varangle ({\vek q},{\vek s})
\end{eqnarray}
are used:
\begin{equation}\label{c6}
X_2=\int\limits_1^\infty dq\int\limits_0^1 dp\ \frac{16\ \pi}{q^2-p^2}\int\limits_{-1}^{+1}\
\frac{dx}{{\bar a}^2-x^2}\int\limits_{s\geq \bar\alpha,\ \cos \varphi_q>0}\frac{sds\ d\varphi_q}{1+s^2}\ \frac{1}{2sq}\ 
\delta\left(\frac{\bar\alpha}{s}-y\right)\ \Theta(\bar\beta-|t|)\ .
\end{equation}
$t$ is introduced to replace $s$ after having done the $\varphi_q$ integration. To this purpose
the relation between $t$ and $s$ is needed. It follows from
$\varphi_q+\varphi_p=\varphi$ or $\varphi_p=\varphi-\varphi_q$, therefore
$\cos \varphi_p=\cos (\varphi-\varphi_q)=
\cos \varphi\cos \varphi_q+\sin \varphi \sin \varphi_q$, what is in terms of $t/s$, $x$, and
$y$ (which equals $+\bar\alpha/s$ because of the delta function):
\begin{equation}\label{c7}
\frac{t}{s}=x\frac{\bar\alpha}{s}\pm\sqrt {1-x^2}\sqrt {1-\left(\frac{\bar\alpha}{s}\right)^2}
\quad {\rm or} \quad (t-\bar\alpha x)^2=(1-x^2)(s^2-\bar\alpha^2)\ .
\end{equation}
This has the consequences
\begin{eqnarray}\label{c8}
s^2-\bar\alpha^2=\frac{(t-\bar\alpha x)^2}{1-x^2}\ , \\
\frac{s ds}{1+s^2}=\frac{(t-\bar\alpha x)dt}{(t-\bar\alpha x)^2+(1+\bar\alpha^2)(1-x^2)}\ .
\end{eqnarray}
With the help of Eq. (\ref{c8}) the $\varphi_q$ integration yields a function depending on $s$
(respectively on $t$). Note that $\cos \varphi_q=+\frac{\bar\alpha}{s}$ has two solutions 
$\varphi_{q,1}$ and
$\varphi_{q,2}$ with $|\sin \varphi_{q,1}|=|\sin \varphi_{q,2}|=\sqrt {1-(\frac{\bar\alpha}{s})^2}$ :
\begin{eqnarray}\label{c10}
\int\limits_0^{2\pi}\frac{d\varphi_q}{s}\ \left[\delta(+\sin\varphi_{q,1} \cdot (\varphi_q-\varphi_{q,1}))+
\delta(+\sin\varphi_{q,2} \cdot (\varphi_q-\varphi_{q,2}))\right] \nonumber \\
= \int\limits_0^{2\pi}\frac{d\varphi_q}{s}\ 
\frac{\delta(\varphi_q-\varphi_{q,1})+\delta(\varphi_q-\varphi_{q,2})}{\sqrt {1-(\frac{\bar\alpha}{s})^2}}=
\frac{2}{\sqrt {s^2-\bar\alpha^2}}=2\ \frac{\sqrt {1-x^2}}{|t-\bar\alpha x|}\ .
\end{eqnarray}
Thus
\begin{eqnarray}\label{c11}
\int\limits_{s\geq \bar\alpha,\ \cos \varphi_q>0}\frac{sds\ d\varphi_q}{1+s^2}\ \frac{1}{2sq}\
\delta\left(\frac{\bar\alpha}{s}-y\right)\ \Theta(\beta-|t|)&=&
\frac{\sqrt {1-x^2}}{q}\int\limits_{t_0}^{t_1}\frac{{\rm sign}(t-\bar\alpha x)\ dt}{(t-\bar\alpha x)^2+
(1+\bar\alpha^2)(1-x^2)} \nonumber \\
&=&\left.\frac{2}{1+q^2}\arctan \frac{t-\bar\alpha x}{\sqrt {(1+\bar\alpha^2)(1-x^2)}}\right |_{t_0}^{t_1}.
\end{eqnarray}
$|t|\leq \bar\beta$ contains the upper limit of the $t$-integration as $t\leq \bar\beta=:t_1$. What 
concerns its lower limit $t_0$, the relation $t\geq -\bar\beta$ competes with $t\geq \bar\alpha x$, 
as it follows from Eq. (\ref{c7}) for the lower limit $\bar\alpha$ of the $s$-integration. This means 
for the ranges of the $t$- and $x$-integrations, one has in the (horizontal) stripe $q=1\cdots\infty,
p=0\cdots1$ of the $q$-$p$-plane to distinguish between the two regions, cf. Fig. 4: \\
(a) region A with $qp\geq 1$ or ${\bar\alpha}\geq {\bar\beta}$ or $-{\bar\alpha}\leq -{\bar\beta}$, and \\
(b) region B with $qp\leq 1$ or ${\bar\alpha}\leq {\bar\beta}$ or ${\bar\alpha}x\leq {\bar\alpha}\leq 
{\bar\beta }$, hence $t_0={\bar\alpha}x$. \\
In the case (a) one has again to distinguish between \\
(i) $x=-1 \cdots {-\bar\beta}/{\bar\alpha}$, with the consequence $t_0={-\bar\beta} $ and \\
(ii) $x=-{\bar\beta}/{\bar\alpha}\cdots +{\bar\beta}/{\bar\alpha}$, with the consequence 
$t_0={\bar\alpha}x$. \\
Thus it is
\begin{eqnarray}\label{c12}
\int\limits_{-1}^{+1}dx\int\limits_{t_0}^{+{\bar\beta}}dt=\left\{
\begin{array} {ll}
\int\limits_{-1}^{-{\bar\beta}/{\bar\alpha}}dx\int\limits_{-{\bar\beta} }^{+{\bar\beta}}dt+
\int\limits_{-{\bar\beta}/{\bar\alpha}}^{+{\bar\beta}/{\bar\alpha}}dx\int\limits_{{\bar\alpha}x}
^{+{\bar\beta}}dt \quad {\rm for} \quad qp\geq 1\; {\rm or}\; \bar\alpha\geq \bar\beta\ ,  \\
\int\limits_{-1}^{+1}dx\int\limits_{{\bar\alpha} x}^{+{\bar\beta}}dt \; 
\quad \qquad \qquad \qquad {\rm for}
\quad qp\leq 1\; {\rm or }\; \bar\alpha\leq \bar\beta\ .
\end{array}
\right.
\end{eqnarray}
With the abbreviations (\ref{c5}) and with
\begin{equation}\label{c13}
\bar f(t,x)= \frac{2}{\bar a^2-x^2}\ \arctan \frac{t-\bar\alpha x}{\sqrt {(1+\bar\alpha^2)(1-x^2)}}\ ,
\end{equation}
Eq. (\ref{c6}) can be written as (note that $\bar f(\bar\alpha x,x)=0$)
\begin{eqnarray}\label{c14}
X_2=\int\limits_1^\infty dq \int\limits_0^1 dp\ \frac{16\ \pi}{q^2-p^2}\ \frac{1}{1+q^2}\
\left\{\Theta(1-qp) \int\limits_{-1}^{+1}dx\ \bar f(\bar\beta,x)\ \right. \nonumber \\
+\left.\Theta(qp-1)\left[\int\limits_{-1}^{-\bar\beta/\bar\alpha}dx\ [\bar f(\bar\beta,x)-\bar f(-\bar\beta,x)]+
\int\limits_{-\bar\beta/\bar\alpha}^{+\bar\beta/\bar\alpha}dx\ \bar f(\bar\beta,x) \right]\right\}\ .
\end{eqnarray}
With the identity
\begin{equation}\label{c15}
\int\limits_{-1}^{-\bar\beta/\bar\alpha}dx\ (-1){\bar f}(-\bar\beta,x)=
\int\limits_{-1}^{-\bar\beta/\bar\alpha}dx\ {\bar f}(\bar\beta,-x)=
\int\limits_{+\bar\beta/\bar\alpha}^{+1}dx\ {\bar f}(\bar\beta,x)
\end{equation}
the terms for $qp\geq 1$ can be comprised as $\int\limits_{-1}^{+1}dx\ {\bar f}(\bar\beta,x)$. Therefore
\begin{equation}\label{c16}
X_2=\int\limits_1^\infty dq\int\limits_0^1 dp\int\limits_{-1}^{+1}dx\
\frac{16\ \pi}{q^2-p^2}\ \frac{1}{1+q^2}\ {\bar f}(\bar\beta,x)\ .
\end{equation}
This is similar to Eq. (\ref{b16}), but there are also differences.
The substitution $q=1/k$ transforms the region of the last two integrations from the 
(horizontal) stripe $q=1\cdots \infty, p=0\cdots 1$ to the more simple unit square $k=0\cdots 1, 
p=0\cdots 1$. With the abbrevations
\begin{equation}\label{c17}
\bar\alpha=\frac{1-k^2}{2k},\quad \bar\beta=\frac{1-p^2}{2p}, \quad \bar a=\frac{1+k^2p^2}{2kp}
\end{equation} 
it is
\begin{eqnarray}
X_2=\int\limits_0^1 dk\int\limits_0^1 dp\int\limits_{-1}^{+1} dx\ 
\frac{16\ \pi}{1-k^2p^2}\ \frac{k^2}{1+k^2}\ {\bar f}(\bar\beta,x)\ . \nonumber
\end{eqnarray} 
Changing finally the notation with $k\to q$ and $p\to k$ makes $\bar\alpha=\alpha$, 
$\bar\beta=\beta$, and
${\bar a}=a$, cf. Eq. (\ref{b17}). With these identities and with $\bar f(t,-x)=f(t,x)$ a further rewriting yields
\begin{equation}\label{c18}
X_2=\int\limits_0^1 dq\int\limits_0^1 dk \int\limits_{-1}^{+1} dx\ 
\frac{16\ \pi}{1-q^2k^2}\ \frac{q^2}{1+q^2}\ f(\beta,x)\approx 21.284906\cdots\ . 
\end{equation} 
This integral differs from Eq. (\ref{b18}) 'only' in an additional factor of $-q^2$ in the nominator of the integrand.
Mathematica5.2 \cite{math} yields the given figure. It seems to hold
\begin{equation}\label{c19}
X_2=\pi^4\left[\frac{2}{3}\ln 2-2\ \frac{\zeta(3)}{\pi^2}\right]\approx 21.284905670516334\cdots.
\end{equation}
How to derive this analytically ?  Is this possible with the method of ref. \cite{Gla2} ? \\

\section{The calculation of $X$}
\setcounter{equation}{0}
\noindent
With Eqs. (\ref{b13}), (\ref{b18}), (\ref{c18}), and with $a$, $\alpha$, $\beta$ being defined in Eq. (\ref{b17}) 
the result for $X=X_1+X_2$ is
\begin{equation}\label{d1}
X=-\int\limits_0^1 dq \int\limits_0^1 dk \int\limits_{-1}^{+1}dx\ 
\frac{16\ \pi}{1-q^2k^2}\ \frac{1-q^2}{1+q^2}\frac{2}{a^2-x^2}\
\arctan \frac{\beta+\alpha x}{\sqrt {(1+\alpha^2)(1-x^2)}}\approx -9.42108\cdots\ .
\end{equation} 
It seems to hold \cite{foo,Gla2}
\begin{equation}\label{d2}
X=-\pi^4\left[\frac{2}{3}\ln 2- 3\ \frac{\zeta(3)}{\pi^2}\right]\approx -9.421079568732553 \cdots.
\end{equation}
The final result \cite{Shi}
\begin{equation}\label{d3}
\Sigma_{2{\rm x}}=-\frac{(\alpha r_s)^2}{4\pi^4}X=e_{2{\rm x}}
\quad {\rm with} \quad \frac{e_{2{\rm x}}}{(\alpha r_s)^2}=\frac{1}{6}\ln 2-\frac{3}{4}\frac{\zeta(3)}{\pi^2}\approx 0.0241792
\end{equation}
appears in the weak-correlation limit of the Hugenholtz-van Hove (Luttinger-Ward) theorem \cite{Hug,Zie4}. 
Because of $\mu_{2{\rm x}}=e_{2{\rm x}}$ it holds the sum rule $\mu_{2{\rm x}}=\Sigma_{2{\rm x}}$ analogous to $\mu_{\rm x}=
\Sigma_{\rm x}$.
Whether perhaps also the more general expression $\Sigma_{2{\rm x}}(k,\omega)$ can be calculated in a similar 
way, has to be studied.

\section*{Acknowledgments}
\noindent
The author thanks M.L. Glasser, U. von Barth, G. R\"opke, and R. Zimmermann for valuable hints, U. Saalmann for 
checking the integrations of 
Eqs. (\ref{b18})  and (\ref{c18}) with Mathematica5.2 \cite{math} by independent Fortran calculations, E. Mrosan 
for a very carefully critical reading the manuscript, F. Tasnadi for his help,  
and acknowledges P. Fulde for supporting this work.

\begin{appendix}
\section*{Appendix A: Derivation of $\Sigma_{2{\rm x}}$ }
\setcounter{equation}{0}
\renewcommand{\theequation}{A.\arabic{equation}}
\noindent
The one-body Green's function of the non-interacting system (ideal Fermi gas)
\begin{equation}\label{A1}
G_0(k,\omega) =\frac{\Theta(k-1)}{\omega-\frac{1}{2}k^2+{\rm i}\delta}+
\frac{\Theta(1-k)}{\omega-\frac{1}{2}k^2-{\rm i}\delta}
\end{equation}
(with $\Theta(x)$ = Heaviside step function) and $G(k,\omega)$, the one-body Green's function of the fully 
interacting system, define the self-energy 
$\Sigma(k,\omega)$ through
\begin{equation}\label{A2}
G(k,\omega)=G_0(k,\omega)+G_0(k,\omega)\Sigma(k,\omega)G(k,\omega)\ . 
\end{equation} 
$\Sigma(k,\omega)$ appears in the Hugenholtz-van Hove theorem (in the Luttinger-Ward form $\mu-\mu_0=\Sigma(1,\mu)$
with $\mu=$ chemical potential)  \cite{Hug} and in the Galitskii-Migdal formula \cite{Gal}
\begin{equation}\label{A3}
v=\frac{1}{2}\int d(k)^3 \int \frac{d\omega}{2\pi{\rm i}}{\rm e}^{{\rm i}\omega\delta}G(k,\omega)\Sigma(k,\omega)\ ,
\quad \delta{_> \atop ^{\to}}0\ .
\end{equation} 
$v$ is the potential component of $e$, the total energy per particle. The contour of the $\omega$-integration is to 
be closed in the upper complex $\omega$-plane. In lowest order it is $\Sigma_{\rm x}(k)=-(1+\frac{1-k^2}{2k}
\ln|\frac{1+k}{1-k}|)$. This makes $v_{\rm x}=-\frac{3}{4}\frac{\alpha r_s}{\pi}$, in agreement with 
$v_{\rm x}=e_{\rm x}$, what follows from the virial theorem $v=r_s\frac{d}{dr_s}e$. \\ 

\noindent
From the Feynman diagram for the exchange term of the 2nd-order self-energy it follows
\begin{eqnarray}\label{A4}
\Sigma_{2{\rm x}}(k,\omega)&=&\frac{(\alpha r_s)^2}{4\pi^4}\int\frac{d^3q_1d^3q_2}{q_1^2q_2^2}\int
\frac{d\eta_1d\eta_2}{(2\pi{\rm i})^2}\times  \\
&\times& G_0(|{\mbox{\bm $k$}}+{\mbox{\bm $q$}}_2|,\omega+\eta_2)G_0(|{\mbox{\bm $k$}}+
{\mbox{\bm $q$}}_1+{\mbox{\bm $q$}}_2|,\omega+\eta_1+\eta_2)G_0(|{\mbox{\bm $k$}}+{\mbox{\bm $q$}}_1|,\omega+\eta_1)
\nonumber 
\end{eqnarray}
Use of (\ref{A1}) yields
\begin{eqnarray}\label{A5}
\Sigma_{2{\rm x}}(k,\omega)=-\frac{(\alpha r_s)^2}{4\pi^4}\int\frac{d^3q_1d^3q_2}{q_1^2q_2^2}
&\left[\D\frac{\Theta(|{\mbox{\bm $k$}}+{\mbox{\bm $q$}}_1+{\mbox{\bm $q$}}_2|-1)
\Theta(1-|{\mbox{\bm $k$} }+{\mbox{\bm $q$}}_1|)\Theta(1-|{\mbox{\bm $k$}
}+{\mbox{\bm $q$}}_2|)}{\omega-\frac{1}{2}k^2+{\mbox{\bm $q$}}_1\cdot{\mbox{\bm $q$}}_2-
{\rm i} \delta}\right. & \nonumber \\
&+\left.\D\frac{\Theta(1-|{\mbox{\bm $k$}}+{\mbox{\bm $q$}}_1+{\mbox{\bm $q$}}_2|)
\Theta(|{\mbox{\bm $k$} }+{\mbox{\bm $q$}}_1|-1)\Theta(|{\mbox{\bm $k$}
}+{\mbox{\bm $q$}}_2|-1)}{\omega-\frac{1}{2}k^2+{\mbox{\bm $q$}}_1\cdot{\mbox{\bm $q$}}_2+
{\rm i} \delta}\right]& . \nonumber \\
\end{eqnarray}
One may check this expression by using it in the Galitskii-Migdal formula (\ref{A3}). Its lhs is known from the virial
theorem as $v_{2{\rm x}}=2e_{2{\rm x}}$ with $e_{2{\rm x}}=$ energy in second order of exchange, calculated by Onsager 
et al. \cite{Ons}. Its rhs gives with Eqs. (\ref{A1}) and (\ref{A5})
\begin{eqnarray}\label{A6}
{\rm rhs}= -\frac{3(\alpha r_s)^2}{(2\pi)^5}\left[\int\frac{d^3kd^3q_1d^3q_2}{q_1^2q_2^2}
\frac{\Theta(1-k)\Theta(1-|{\vek k}+{\vek q}_1+{\vek q}_2|)
\Theta(|{\vek k}+{\vek q}_1|-1)\Theta(|{\vek k}+{\vek q}_2|-1)}{{\vek q}_1\cdot{\vek q}_2+{\rm i}\delta}\right.
\nonumber \\
\left.+\int\frac{d^3kd^3q_1d^3q_2}{q_1^2q_2^2}\frac{\Theta(k-1)\Theta(|{\vek k}+{\vek q}_1+{\vek q}_2|-1)
\Theta(1-|{\vek k}+{\vek q}_1|-1)\Theta(1-|{\vek k}+{\vek q}_2|)}{{\vek q}_1\cdot(-{\vek q}_2)+{\rm i}\delta}\right]\ .
\nonumber \\
\end{eqnarray}
It is easy to show with the help of the substitutions ${\vek q}_1\to{\vek q}_1'$, ${\vek q}_2\to -{\vek q}_2'$, 
${\vek k}\to -({\vek k}'+{\vek q}_1')$ that the second term equals the first one. Thus
\begin{eqnarray}\label{A7}
{\rm Re} \; {\rm rhs}&=&
-2 \frac{3(\alpha r_s)^2}{(2\pi)^5}\int\frac{d^3kd^3q_1d^3q_2}{q_1^2q_2^2}\frac{P}{{\vek q}_1\cdot{\vek q}_2}\times \\
&\times&\Theta(1-k)\Theta(1-|{\vek k}+{\vek q}_1+{\vek q}_2|)
\Theta(|{\vek k}+{\vek q}_1|-1)\Theta(|{\vek k}+{\vek q}_2|-1)=2e_{2{\rm x}}\ . \nonumber
\end{eqnarray} 
$P$ means the Cauchy principle value. This is in agreement with the above mentioned relation.  That $e_{2{\rm x}}$ of 
Eq. (\ref{A7}) really agrees with the integral calculated by Onsager et al. \cite{Ons} follows from the substitutions
${\vek k}\to{\vek k}_1$, ${\vek q}_1\to {\vek q}$, ${\vek q}_2\to -({\vek k}_1+{\vek k}_2+{\vek q})$. From Eq. 
(\ref{A5}) also follows $n_{2{\rm x}}(k)$, the second-order-in-exchange contribution to the momentum distribution.
It is again easy to derive the well-known asymptotics $n_{2{\rm x}}(k\to\infty)=-\frac{4}{9\pi^2}
\frac{(\alpha r_s)^2}{k^8}$. \\

\noindent
After this control of $\Sigma_{2{\rm x}}(k,\omega)$, the formula for $\Sigma_{2{\rm x}}={\rm Re}\ 
\Sigma_{2{\rm x}}(1,1/2)$ follows from Eq. (\ref{A5}) as 
$\Sigma_{2{\rm x}}= -\frac{(\alpha r_s)^2}{4\pi^4}(X_1+X_2)$, where $X_{1,2}$ mean the integrals 
\begin{eqnarray}\label{A8} 
X_1=\int d^3q_1\ d^3q_2\ \frac{P}{{\vek q}_1 \cdot {\vek q}_2 }\ \frac{1}{q_1^2\ q_2^2}\
\Theta[1-({\mbox{\bm $e$}}+{\mbox{\bm $q$}}_1+{\mbox{\bm $q$}}_2)^2]
\Theta[({\mbox{\bm $e+q$}}_1)^2-1]\Theta[({\mbox{\bm $e+q$}}_2)^2-1]\ , \nonumber \\  
\end{eqnarray}
\begin{eqnarray}\label{A9}
X_2=\int d^3q_1\ d^3q_2\ \frac{P}{{\vek q}_1 \cdot {\vek q}_2 }\ \frac{1}{ q_1^2\ q_2^2}
\Theta[({\mbox{\bm $e$}}+{\mbox{\bm $q$}}_1+{\mbox{\bm $q$}}_2)^2-1]
\Theta[1-({\mbox{\bm $e+q$}}_1)^2]\Theta[1-({\mbox{\bm $e+q$}}_2)^2]\ . \nonumber \\
\end{eqnarray}
They contain ${\vek q}_1\cdot{}{\vek q}_2$ as the energy denominator.
$1/q_{1,2}^2$ arises from the Coulomb repulsion and the remainder is due to the Pauli principle.
${\mbox{\bm $e$}}$ is a unit vector.
Note that ${\vek q}_1\cdot{\vek q}_2<0$ for $X_1$, because of $(2{\vek e}+{\vek q}_1)\cdot{\vek q}_1>0$ and 
$(2{\vek e}+{\vek q}_2)\cdot{\vek q}_2>0$ in combination with $(2{\vek e}+{\vek q}_1)\cdot{\vek q}_1
+(2{\vek e}+{\vek q}_2)\cdot{\vek q}_2+2{\vek q}_1\cdot{\vek q}_2<0$. This latter inequality enforces 
${\vek q}_1\cdot{\vek q}_2<0$. Thus $X_1<0$. It follows similarly $X_2>0$.  
The integrals $X_{1,2}$ do not depend on ${\mbox{\bm $e$}}$. Therefore application of ${\hat O}=\int d^3e/(4\pi)\
2\ \delta({\vek e}^2-1)$ does not change them (notice $2\ \delta({\vek e}^2-1)=\delta(e-1)$). Following Onsager et al. 
\cite{Ons}, $\vek e$ is resolved into its 
components perpendicular to the ${\vek q}_1-{\vek q}_2$-plane ${\vek e}_\bot$, and in the plane ${\vek e}_\|$: 
${\hat O}=\int de_\bot/2\ \int d^2e_\|/(2\pi)\ 2\ \delta(e_\|^2+e_\bot^2-1)$. The 
integration over ${\vek e}_\bot$ may be done immediately by means of a change in scale: ${\vek q}_1={\tilde{\vek q}_1}
{\sqrt {1-e_\bot^2}}$, ${\vek q}_2={\tilde{\vek q}_2}
{\sqrt {1-e_\bot^2}}$, ${\vek e}_\|={\tilde {\vek e}_\|}{\sqrt {1-e_\bot^2}}$. The results are (denoting 
${\tilde {\vek e}_\|}$ as ${\vek e}$ and deleting also all the other tildes for simplicity) 
\begin{eqnarray}\label{A10}
X_1=&&\int d^3q_1\ d^3q_2\ \frac{P}{{\vek q}_1 \cdot {\vek q}_2 }\ \frac{1}{q_1^2\ q_2^2}\int\frac{d^2e}{2\pi}\ 2\ 
\delta({\mbox{\bm $e$}}^2-1)\ \times \nonumber \\
&&\times \Theta[1-({\mbox{\bm $e$}}+{\mbox{\bm $q$}}_1+{\mbox{\bm $q$}}_2)^2]
\Theta[({\mbox{\bm $e+q$}}_1)^2-1]\Theta[({\mbox{\bm $e+q$}}_2)^2-1]\ ,                               
\end{eqnarray}
\begin{eqnarray}\label{A11}
 X_2=\int d^3q_1\ d^3q_2\ \frac{P}{{\vek q}_1 \cdot {\vek q}_2 }\ \frac{1}{ q_1^2\ q_2^2}\ \int\frac{d^2e}{2\pi}\ 2\ 
\delta({\vek e}^2-1)\times \nonumber \\
\times \Theta[({\mbox{\bm $e$}}+{\mbox{\bm $q$}}_1+{\mbox{\bm $q$}}_2)^2-1]
\Theta[1-({\mbox{\bm $e+q$}}_1)^2]\Theta[1-({\mbox{\bm $e+q$}}_2)^2]\ . 
\end{eqnarray}
Whereas the two terms of $\Sigma_{2{\rm x}}(k,\omega)$, cf. (\ref{A6}) each contributes $e_{2{\rm x}}$ to $v_{2{\rm x}}$
(thus $v_{2{\rm x}}=2e_{2{\rm x}}$) as shown above, cf. (\ref{A7}), $X_1$ and $X_2$ contribute different values to
$\Sigma_{2{\rm x}}(1,1/2)$ as shown in Secs. II and III. 
\end{appendix}

\begin{figure}[h!]
\resizebox{40mm}{!}{
\includegraphics{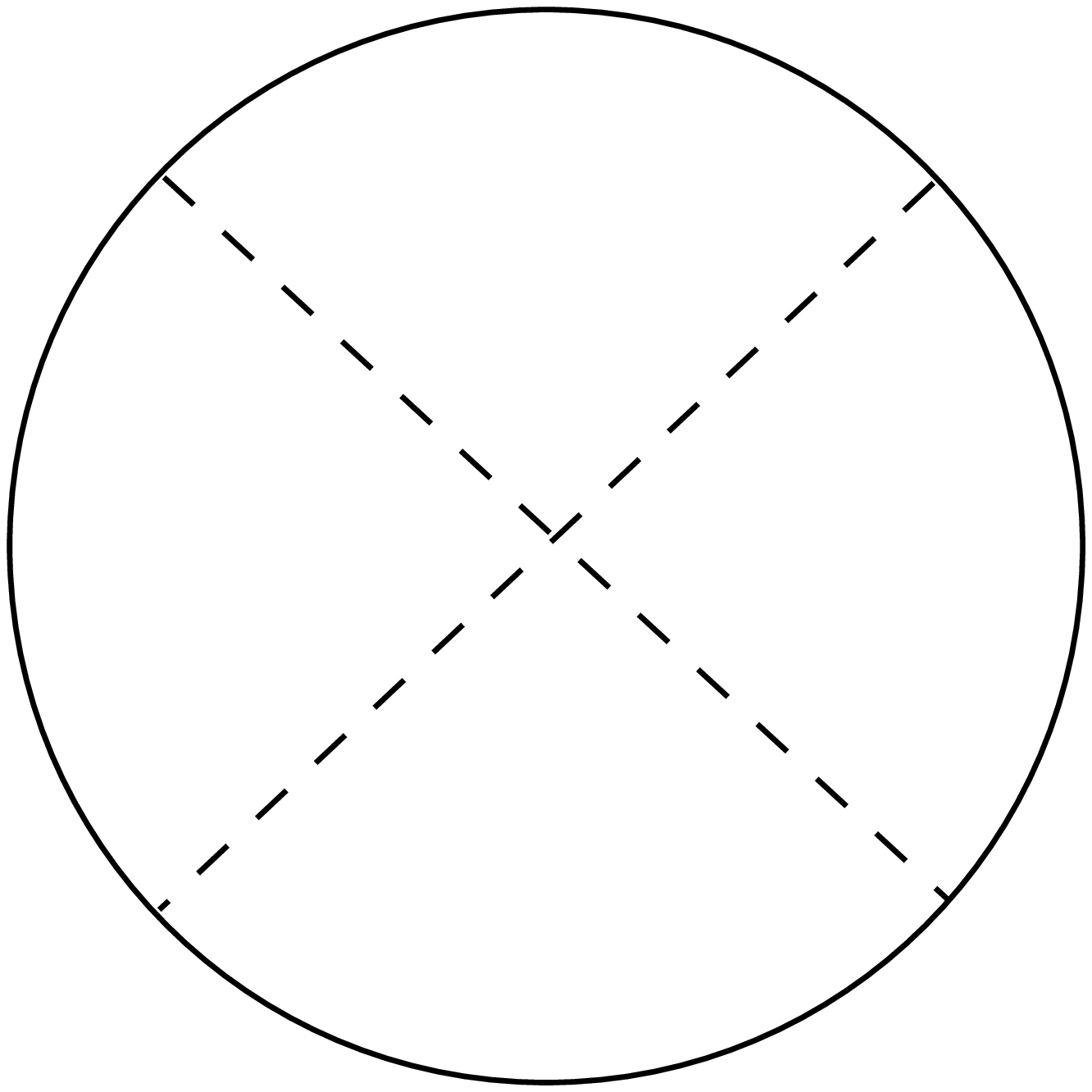}}
\caption{The Feynman diagram of $e_{2{\rm x}}$, analytically calculated by Onsager et al. \cite{Ons}.}
\vspace{2cm}
\resizebox{80mm}{!}{
\includegraphics{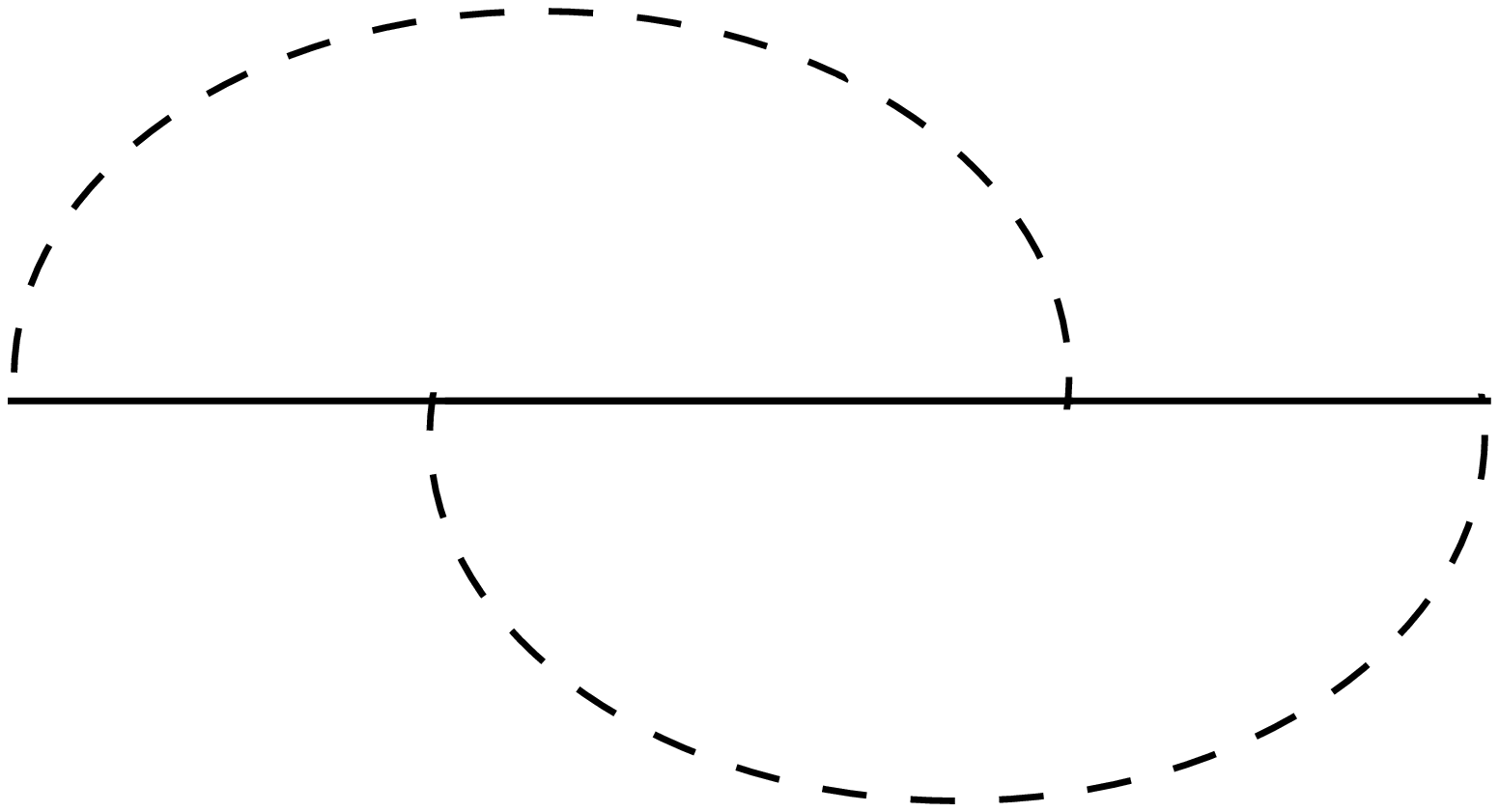}}
\caption{The Feynman diagram of $\Sigma_{2{\rm x}}(k,\omega)$, (semi)analytically calculated in this paper.}
\end{figure}

\begin{figure}[h!]
\resizebox{160mm}{!}{
\includegraphics{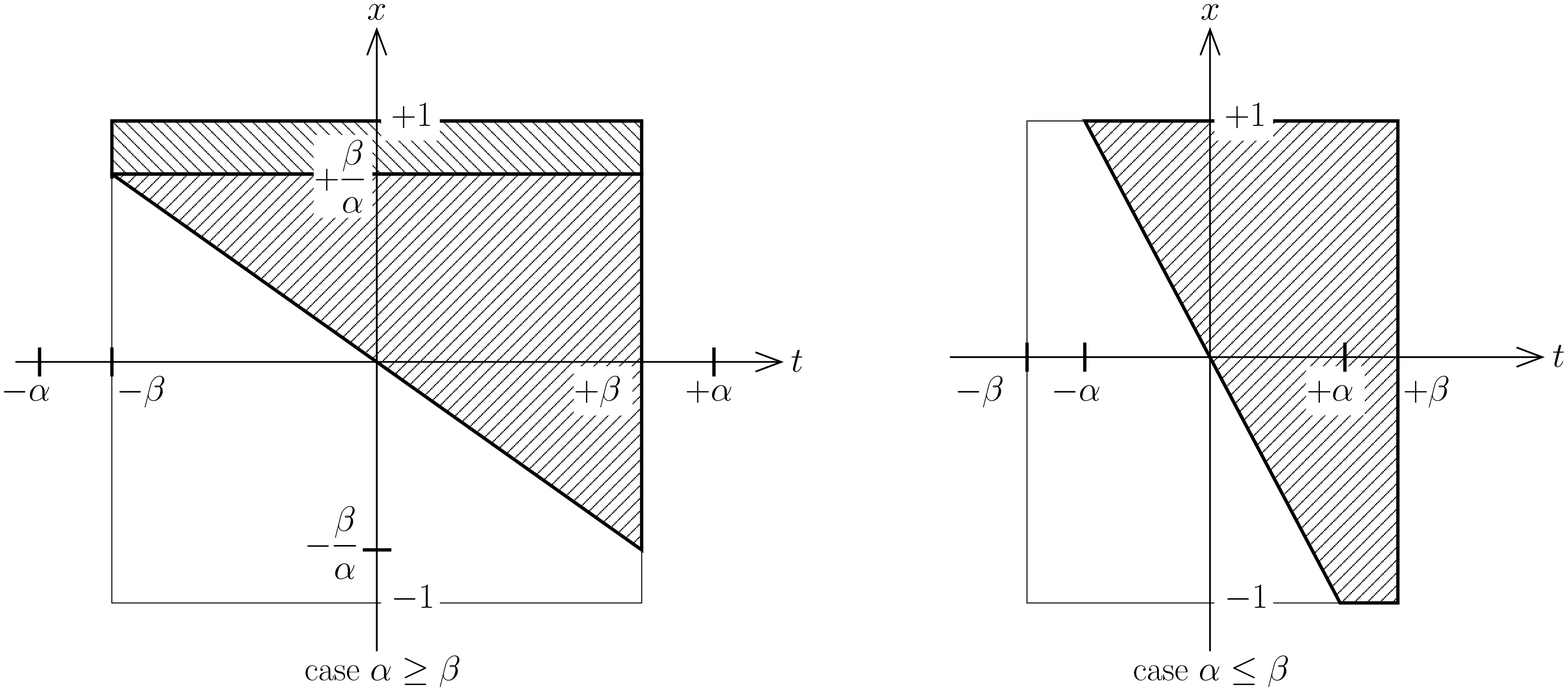}}
\caption{The dashed area is the region of integration described by Eq. (\ref{b12}).}
\vspace{2cm}
\resizebox{160mm}{!}{
\includegraphics{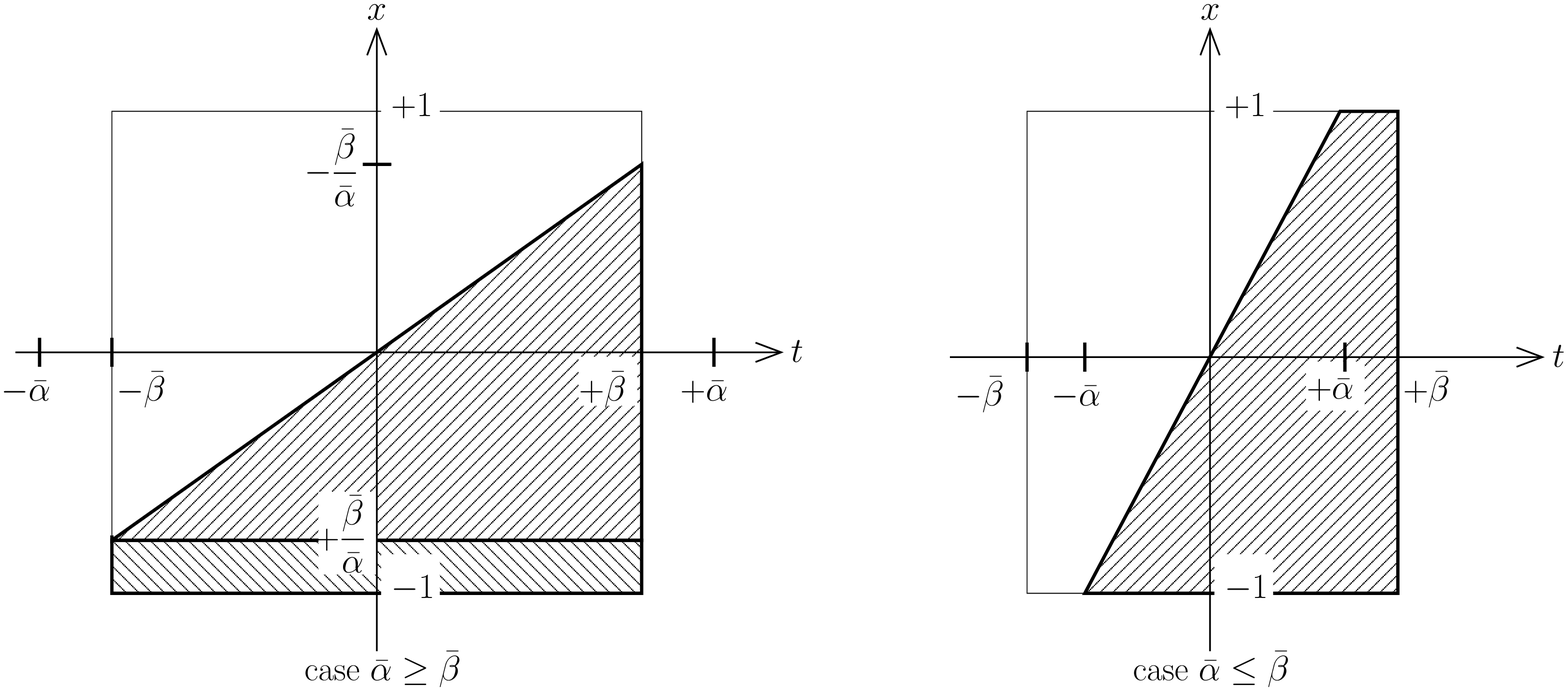}}
\caption{The dashed area is the region of integration described by Eq. (\ref{c12}).}
\end{figure}


\begin{thebibliography}{1}
\bibitem{foo} For $X_1-X_2$ (instead of $X_1+X_2$) it has been shown analytically 
$X_1-X_2=-\pi^4[2\ln 2-7\frac{\zeta(3)}{\pi^2}]$, in agreement with Eqs. (\ref{b19}) and (\ref{c19}) 
\cite{Gla2}.
\bibitem{Shi} This result should be comparable to calculations by E. Shirley \cite{Shir}
if one considers in his Fig. 1 the contribution of the second (exchange)
Feynman diagram to the on-shell self-energy $\Sigma(1,1/2)$ in the limit
$r_s\to 0$, where the effectively screened Coulomb repulsion $W(12)$ is replaced by the bare Coulomb repulsion
$\nu(12)$.
\end{thebibliography}
\end{document}